\documentclass[superscriptaddress,twocolumn]{aastex631}
\usepackage{CJK}
\usepackage{url}
\usepackage{apjfonts}
\usepackage{amsmath,amssymb,amstext}
\usepackage[super]{nth}
\usepackage{microtype}
\usepackage{graphicx}
\usepackage{etoolbox}

\usepackage{hyperref}
\begin{document}



\title{Mapping the Cloud-Driven Atmospheric Dynamics \& Chemistry of an Isolated Exoplanet Analog with Harmonic Signatures}


\shorttitle{Cloud-Driven Atmospheric Dynamics \& Chemistry of an Isolated Exoplanet Analog}

\shortauthors{Plummer, Cocchini, \& Kearns et al.}

\author[0000-0002-4831-0329]{Michael K. Plummer}
\affiliation{Department of Physics and Meteorology, United States Air Force Academy, 2354 Fairchild Drive, CO 80840, USA}

\correspondingauthor{Michael K. Plummer}

\author[0009-0000-4531-3033]{Francis P. Cocchini}
\affiliation{Department of Physics and Meteorology, United States Air Force Academy, 2354 Fairchild Drive, CO 80840, USA}

\author[0009-0000-8040-861X]{Peter A. Kearns}
\affiliation{Department of Physics and Meteorology, United States Air Force Academy, 2354 Fairchild Drive, CO 80840, USA}

\author[0000-0003-2015-5029]{Allison McCarthy}
\affiliation{Department of Astronomy, Boston University, 725 Commonwealth Ave., Boston, MA 02215, USA}
\affiliation{School of Physics, Trinity College Dublin, The University of Dublin, Dublin 2, Ireland}

\author[0000-0003-3506-5667]{\'Etienne Artigau}
\affiliation{D\'epartment de Physique, Universit\'e de Montr\'eal, IREX, Montr\'eal, QC, H3C 3J7, Canada}
\affiliation{Observatoire du Mont-M\'egantic, Universit\'e de Montr\'eal, Montr\'eal, QC, H3C 3J7, Canada}

\author[0000-0001-6129-5699]{Nicolas B. Cowan}
\affiliation{Department of Physics, McGill University, Montreal, Canada
}
\affiliation{Department of Earth \& Planetary Sciences, McGill University, Montreal, Canada}.

\author{Roman Akhmetshyn}
\affiliation{Department of Physics, McGill University, Montreal, Canada
}

\author[0000-0003-0489-1528]{Johanna M. Vos}
\affiliation{School of Physics, Trinity College Dublin, The University of Dublin, Dublin 2, Ireland}

\author[0000-0002-9792-3121]{Evert Nasedkin}
\affiliation{School of Physics, Trinity College Dublin, The University of Dublin, Dublin 2, Ireland}

\author[0000-0001-6627-6067]{Channon Visscher}
\affiliation{Chemistry \& Planetary Sciences, Dordt University, Sioux Center, IA 51250}
\affiliation{Center for Extrasolar Planetary Systems, Space Science Institute, Boulder, CO 80301}

\author[0000-0001-5578-1498]{Bj\"orn Benneke}
\affiliation{D\'epartment de Physique, Universit\'e de Montr\'eal, IREX, Montr\'eal, QC, H3C 3J7, Canada}

\author{Ren\'e Doyon}
\affiliation{D\'epartment de Physique, Universit\'e de Montr\'eal, IREX, Montr\'eal, QC, H3C 3J7, Canada}



\author[0000-0003-3050-8203]{Stanimir A. Metchev}
\affiliation{Western University, Department of Physics and Astronomy, London, Ontario, Canada}
\affiliation{Western University, Institute for Earth and Space Exploration, London, Ontario, Canada}

\author{Jason F. Rowe}
\affiliation{Department of Physics \& Astronomy, Bishop’s University, 2600 Rue College, Sherbrooke, QC J1M 1Z7, Canada}

\author[0000-0002-2011-4924]{Genaro Su\'arez}
\affiliation{Department of Astrophysics, American Museum of Natural History, Central Park West at 79th Street, NY 10024, USA}


\begin{abstract}


Young planetary-mass objects and brown dwarfs near the L/T spectral transition exhibit enhanced spectrophotometric variability over field brown dwarfs. Patchy clouds, auroral processes, stratospheric hot spots, and complex carbon chemistry have all been proposed as potential sources of this variability. Using time-resolved, low-to-mid-resolution spectroscopy collected with the JWST/NIRISS and NIRSpec instruments, we apply harmonic analysis to SIMP J013656.5+093347, a highly variable, young, isolated planetary-mass object. Odd harmonics ($k=3$) at pressure levels ($\gtrsim 1~$bar) corresponding to iron (Fe) and forsterite (Mg$_{2}$SiO$_{4}$) cloud formation suggest potential North/South hemispheric asymmetry in the cloudy, and likely equatorial, regions. We use the inferred harmonics, along with 1-D substellar atmospheric models, to map the flux variability by atmospheric pressure level. These vertical maps demonstrate robust interaction between deep convective weather layers and the overlying stratified and radiative atmosphere. We identify distinct time-varying structures in the near-infrared that we interpret as planetary-scale wave (e.g., Rossby or Kelvin)-associated cloud modulation. We detect deviations from bulk (composite) variability in water ($\rm{S/N_{max}=14.0}$), carbon monoxide ($\rm{S/N_{max}=13.0}$), and methane ($\rm{S/N_{max}=14.9}$) molecular signatures. Forsterite cloud modulation is anti-correlated with overlying carbon monoxide and water abundances and correlated with deep methane absorption, suggesting complex interaction between cloud formation, atmospheric chemistry, and temperature structure. Furthermore, we identify distinct harmonic behavior between methane and carbon monoxide absorption bands, providing evidence for time-resolved disequilibrium carbon chemistry. At the lowest pressures ($\lesssim100~$mbar), we find that the mapped methane lines transition from absorption to emission, supporting evidence of high-altitude auroral heating via electron precipitation.

\end{abstract}


\section{Introduction}\label{sec:Intro}

Highly variable and rapidly rotating planetary-mass objects and low-mass brown dwarfs provide the opportunity to efficiently observe dynamic gas giant exoplanet-analog atmospheres over multiple rotations, unhindered by host star contamination \citep{Burrows2001,Faherty2016}. These objects, with inferred masses ($\lesssim 15~ M_{J}$) overlapping the lowest mass products of stellar formation and the highest mass products of planetary formation, serve as laboratories for understanding how atmospheric circulation, cloud formation, molecular chemistry, and auroral processes interact in H/He-dominated planetary environments.

Brown dwarfs at the transition ($\rm{T_{eff}} \sim 1300~$K) from mineral cloud-rich late-L dwarfs to the more cloud-sparse mid-T dwarfs exhibit amplified variability \citep{Radigan2014a,Radigan2014b,Eriksson2019,Liu2024}. Driving sources of this variability are believed to include multiple, patchy cloud layers perhaps of varying thickness \citep[e.g.,][]{ackerman&marley01,Burgasser2002,Reiners&Basri2008,Apai2013}; planetary-scale waves \citep[for e.g., Rossby or Kelvin waves,][]{Apai2017,Apai2021,Zhou2020b,Fuda:2024}; cloud modulation driven by those planetary-scale waves \citep{Plummer2024}; double-diffusive convection tied to disequilibrium carbon chemistry \citep{Tremblin2015, Tremblin2016}; upper atmospheric hot spots \citep{Robinson&Marley2014,Morley2014}; dynamically driven temperature variations \citep{showman&kaspi13,Nasedkin2025}; and potentially auroral processes \cite[][]{Hallinan2015,Kao2016,Kao2018,McCarthy2025,Nasedkin2025,Suarez2025}.
\subsection{SIMP~J0136}

\par SIMP J013656.5+093347.3 (hereafter SIMP~J0136) \citep{Artigau2006} is an L/T transition object (T$2.5 \pm 0.5$) that is highly variable ($\gtrsim 2\%$) at near-infrared (NIR) wavelengths \citep[see e.g.,][]{Artigau2009}. With a mass \citep[$12.7 \pm 1.0$ M$_{J}$,][]{Gagne2017} straddling the minimum deuterium-burning limit, it is often referred to as a planetary-mass object. A relatively young object \citep[$200 \pm 50$ Myr,][]{Gagne2017} with an inferred surface gravity $\rm{\log(g)} = 4.5 \pm 0.4$ \citep{Vos2023}, SIMP~J0136 fits into a population of younger and lower surface gravity brown dwarfs exhibiting enhanced variability rates when compared to field brown dwarfs \citep{Vos2022,Liu2024}. SIMP~J0136 has a rapid rotation rate \citep[$2.414 \pm 0.078$ h,][]{Yang2016}, likely contributing to amplified atmospheric dynamics \citep[e.g., increased wind speeds,][]{Allers2020}, larger equator-to-pole cloud contrasts due to Coriolis effects \citep{tan21a,tan21b}, and auroral emission due to a strengthened magnetic dynamo and associated currents \citep[e.g.,][]{Kao2018}.

\par Multi-modal observations suggest SIMP~J0136's atmosphere is structured into multiple, time-varying mineral cloud layers. \texttt{Brewster} framework \citep{Burningham2017} retrievals on spectra spanning $1-15~\mu$m \citep{Burgasser2008,Sorahana2013,Filippazzo2015} infer a thin and patchy forsterite (Mg$_2$SiO$_4$) cloud layer overlying a thicker and more spatially expansive iron (Fe) cloud deck \citep{Vos2023}. Ground-based photometry has detected phase shifts between the NIR bands, supporting $\geq 2$ cloud layers \citep{Yang2016,McCarthy:2024,Plummer2024}. The suspected forsterite clouds’ phase (inferred via color index reddening) anti-correlation with variation in broadband NIR emission suggest planetary-scale waves may be related to cloud formation and dissipation \citep{Plummer2024}, similar to ammonia/ammonium hydrosulfide cloud modulation in Jupiter’s atmosphere \citep[e.g.,][]{Choi2013,dePater:2016,Fletcher:2016,Fletcher2020}. 

\par With its ability to collect spectroscopic and photometric observations across the NIR and mid-IR spectrum, the James Webb Space Telescope (JWST) has revolutionized our understanding of substellar atmospheres like SIMP~J0136. Applying a K-means cluster algorithm to time-resolved $0.8-11~\mu$m spectra resulted in 9 characteristic lightcurve shapes (or clusters) for JWST NIR Spectrograph \citep[NIRSpec,][]{jakobsen_nirspec_2023} data and 2 characteristic lightcurves for JWST Mid-IR Instrument \citep[MIRI,][]{wright_miri_2023} data \citep{McCarthy2025}. These characteristic lightcurves potentially correspond to distinct atmospheric layers \citep[i.e., iron and forsterite cloud layers and cloud-free regions,][]{McCarthy2025}. The K-means algorithm was previously applied to the WISE~1049\,AB (L7.5/T0.5) binary brown dwarf system, finding 3 characteristic and spectrophotometrically-resolved lightcurve shapes for the NIRSpec data and 2 characteristic lightcurves for the MIRI data \citep{Biller2024}. A follow-on epoch of observations of WISE~1049\,AB with the same instruments found similar results and tied the variation seen in the characteristic lightcurves to patchy clouds and higher-altitude hot spots \citep{Chen2025}. 

\par The JWST Near Infrared Imager and Slitless Spectrograph (NIRISS; \citealt{doyon_niriss_2023}) instrument has also delivered important findings on SIMP~J0136. Principal Component Analysis of NIRISS time-series spectroscopy found that rotational variability was driven by $\geq3$ spectrally distinct regions \citep{Akhmetshyn2025}, broadly agreeing with \citet{McCarthy2025}’s findings. Multi-wavelength lightcurves collected by NIRISS identified odd harmonics, suggesting asymmetry between Northern and Southern hemispheres, and found initial evidence that broadband emission modulation from forsterite spectral regions was anti-correlated with variability in water and CO absorption lines \citep{Akhmetshyn2025}.

\par SIMP~J0136 possesses a robust magnetic field potentially capable of strong auroral emissions. Radio observations with the Karl G. Jansky Very Large Array at 8-12 GHz detected a $\sim 3000~$G magnetic field \citep{Kao2018}, much stronger than Jupiter's $\sim4~$G equatorial dipole field \citep[see e.g.,][]{Barrow1960,Acuna&Ness1975,Connerney2018,Connerney2022,Sharan2022}. Auroral heating signatures, in the form of H$_3+$ emission at $\sim3~\mu$m, have been detected in NIR observations of Jupiter's upper atmosphere \citep[see e.g.,][]{Kim1993}, but direct detection of H$_3+$ emission remains elusive for T dwarfs like SIMP~J0136, and as discussed in \citet{Nasedkin2025}, may remain so as H$_3+$’s quick reaction rate with water may prevent observable H$_3+$ emission \citep{Pineda2024}. Providing further support for \citet{Pineda2024}'s findings, recent observations of the isolated Y dwarf, CWISEP J193518.59-154620.3 (WISE~J1935), found emission in the $\nu_{3}$ rotational-vibrational methane band \citep{Faherty2024}, suggesting upper atmospheric auroral heating, although H$_3+$ emission was not detected. 

Indirect detection methods for inferring the presence of auroral heating are promising. Time-resolved spectral retrievals \citep[via \texttt{petitRADTRANS,}][]{Molliere2019,Nasedkin2024} of the same JWST/NIRSpec data presented originally in \citet{McCarthy2025} identify a high-altitude, $265\,$K temperature inversion, attributed to tentative auroral heating from electron precipitation \citep{Nasedkin2025}. 

\subsection{Outline}

\par In this work, we infer and vertically map the atmospheric harmonics in the time-resolved NIR spectra of SIMP~J0136. We analyze JWST NIRISS and NIRSpec datasets, collected $\sim 35~$h apart from one another and originally presented in \citet{Akhmetshyn2025} and \citet{McCarthy2025} respectively, and identify connections between cloud modulation and temperature and chemical structure. The article is structured as follows: \S \ref{sec:Observations} describes the observational strategy and data reduction. \S \ref{sec:Models} describes our harmonic model and best-fit scheme. \S \ref{sec:Results} presents the inferred spectral harmonic parameters and uses the best-fit lightcurves and atmospheric models to create vertical (pressure vs. time) heat maps showing the evolution of the atmosphere of SIMP~J0136 as well as spectral features' deviations from bulk atmospheric variability. \S \ref{sec:Discussion} discusses the implications of our results for the dynamical, temperature, and chemical structure of SIMP~J0136 and \S \ref{sec:Summary} summarizes our findings.

\section{Observations and Data Reduction} \label{sec:Observations} 

\par To understand how the atmosphere of SIMP~J0136 evolves in the NIR over time, we consider two epochs of observations by two JWST instruments, NIRISS and NIRSpec. NIRISS collections (GTO 1209, PI: Artigau) being made on Jul 22, 2023 from UT 05:08:28 to 09:05:07 \citep[observational cadence: 1.96~min, total observation time: 2.62~h,][]{Akhmetshyn2025}, and NIRSpec observations (GO 3548, PI: Vos) conducted on July 23, 2023 from UT 18:40:56 to 22:05:06 \citep[observational cadence: 1.83~s, total observation time: 2.90~h,][]{McCarthy2025}. 

\par
\begin{figure*}
\centering
\includegraphics[width=1.0\textwidth]{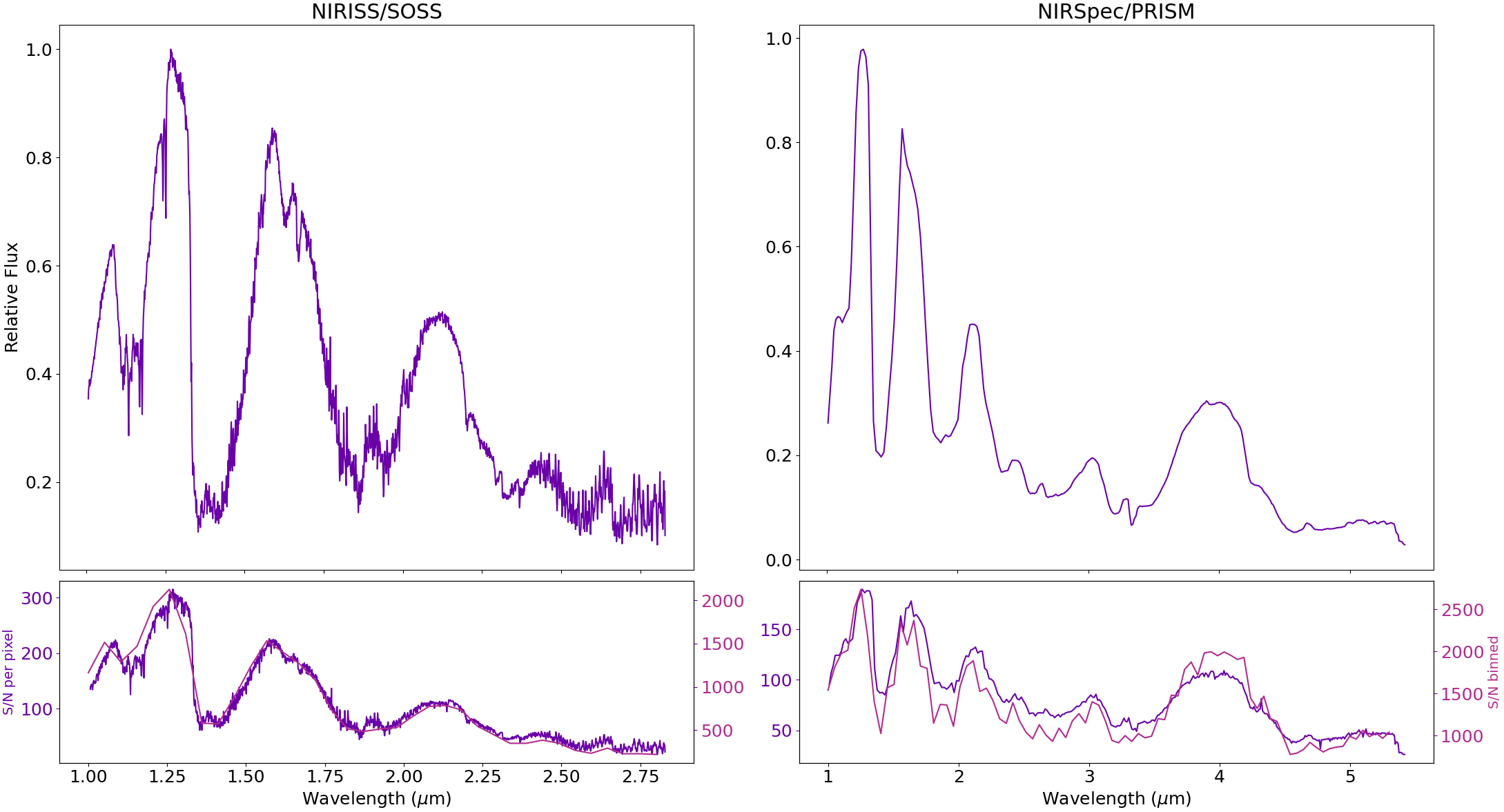}
\caption{\label{fig:spectra} NIRISS/SOSS (left) and NIRSpec/PRISM (right) spectra for SIMP$\,$J0136. \textbf{Top Row:} Time-averaged spectra normalized to maximum emission. \textbf{Bottom Row:} Per pixel (purple) and spectrally and temporally binned (magenta) signal-to-noise ratio. NIRISS/SOSS and NIRSpec/PRISM data were previously presented in \citet{Akhmetshyn2025} and \citet{McCarthy2025,Nasedkin2025}, respectively.
}
\end{figure*}

\begin{figure*}
\centering
\includegraphics[width=1.0\textwidth]{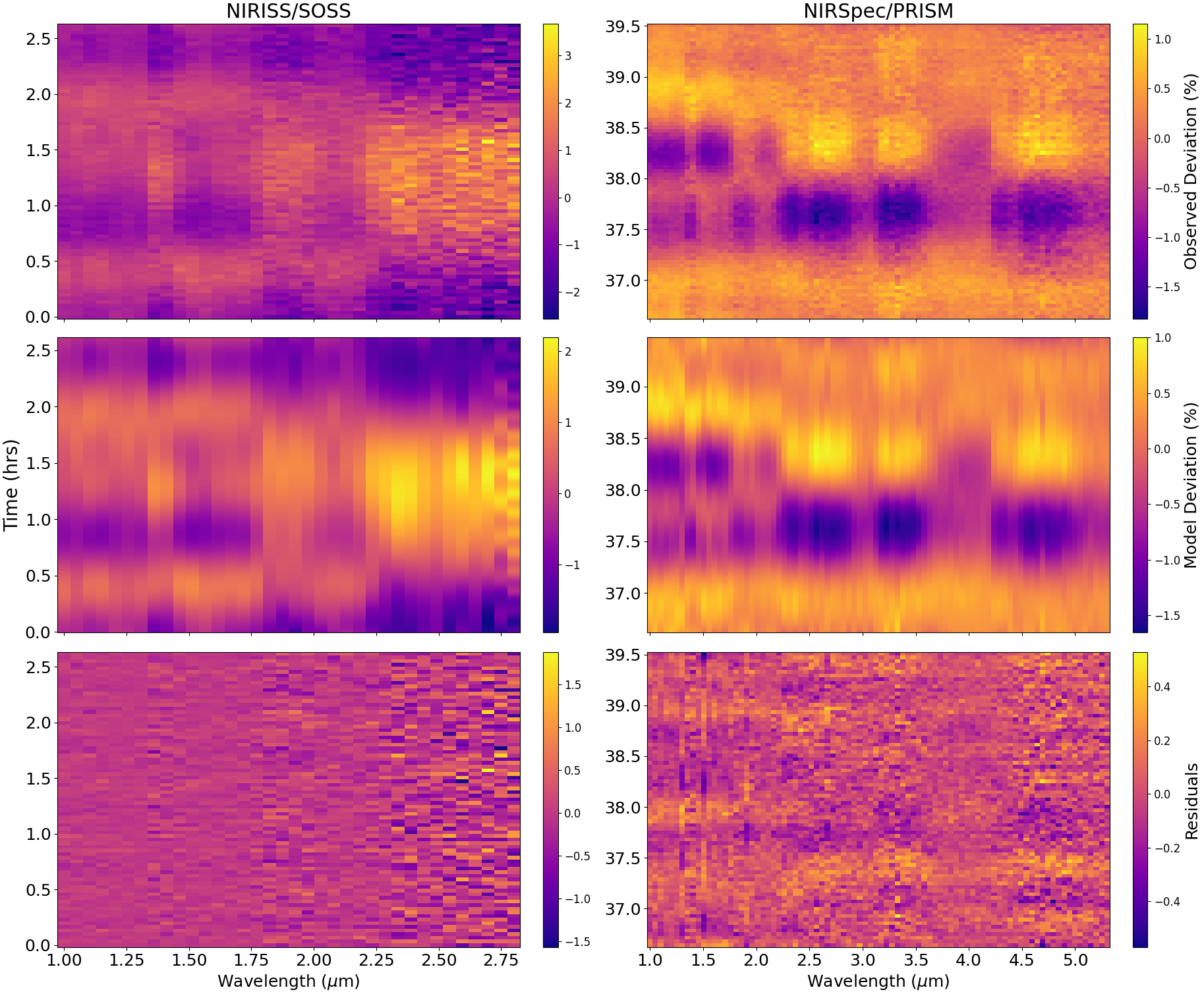}
\caption{\label{fig:dynamic_spectra} NIRISS/SOSS (left) and NIRSpec/PRISM (right) variability maps (dynamics spectra) for SIMP~J0136. Variability maps use 0.05\,$\mu$m spectral binning and demonstrate the normalized spectral variability over the periods of observation. Integration times for each temporal bin are 118\,s (NIRISS/SOSS) and 129\,s (NIRSpec/PRISM). \textbf{Top Row:} Observed variability maps. \textbf{Middle Row:} Best-fit variability maps with lightcurves inferred using harmonic model described in \S \ref{ssec:models:wave_model} and Bayesian nested sampling. \textbf{Bottom Row:} Residuals between observed and best-fit variability maps. NIRISS/SOSS's residuals ($0.00 \pm 0.215$\,\%) approximate white noise while the NIRSpec/PRISM residuals ($0.00 \pm 0.075 \%$) contain seemingly structured harmonics signals of $k \gtrsim 3$, likely due to unmodeled dynamics or systematics. NIRISS/SOSS and NIRSpec/PRISM data were previously presented in \citet{Akhmetshyn2025} and \citet{McCarthy2025,Nasedkin2025}, respectively.
}
\end{figure*}

 Figure~\ref{fig:spectra} includes the mean emission spectra as well as the per pixel and binned (temporally and spectrally) signal-to-noise (S/N) ratio. The NIRSpec/Prism data received a larger increase in binned S/N compared to the NIRISS/SOSS data because of the large number of observations made in NIRSpec's Bright Object Times Series (BOTS) mode. Variability maps (also commonly known as dynamic spectra) are shown in Figure \ref{fig:dynamic_spectra}. The variability maps display the normalized variability at each observed wavelength over the period of observation, which are discussed further in \S \ref{sec:Models} below.

\par Here we will give a brief overview of the observations and data reduction as they were previously presented in detail in \citet{McCarthy2025} and \citet{Akhmetshyn2025}. NIRISS was used in Single Object Slitless Spectroscopy mode (NIRISS/SOSS) to collect 81, $R\sim1200$ spectra from 0.85 to 2.83\,$\mu$m. In this work, we only consider wavelengths $> 1~\mu$m due to background star contamination at shorter wavelengths. The SOSS-Inspired SpectroScopic Extraction (SOSSISSE) pipeline was implemented for data reduction \citep{Lim2023}. NIRSpec BOTS was used in Prism/CLEAR mode to collect 5726, $R = 30-300$ spectra from 0.6 to 5.3\,$\mu$m. As for NIRISS/SOSS above, only wavelengths $>1\,\mu$m are considered. For data reduction, \citet{McCarthy2025} used the JWST Space Telescope Science Institute pipeline version 1.14.0 in default settings.

\par We use $118~$s and $129~$s temporal bins for NIRISS/SOSS and NIRSpec/PRISM respectively to create 81 time steps for each data set. It is worth noting that for the NIRSpec/PRISM dataset, this temporal binning reduces the observed \% deviation by a factor of $\sim2$ versus that seen in \citet{McCarthy2025} and \citet{Nasedkin2025}. Both data sets use $0.05~\mu$m spectral binning to compare spectral features at an equivalent resolution. We bin linearly (rather than logarithmically) to more precisely target specific molecular bands, and we acknowledge that the bins will become fractionally smaller at longer wavelengths.

\section{Models \& Fitting Scheme}\label{sec:Models}

\subsection{Harmonic Model}\label{ssec:models:wave_model}

\par Using the Python code \texttt{Imber} \citep{Imber:2023,Imber:2024} developed in \citet{Plummer&Wang2022,Plummer&Wang2023,Plummer2024}, we fit each lightcurve with models of 2nd and 3rd order Fourier modes (or harmonic components), with and without damping. \texttt{Imber} is an open-source code available on GitHub\footnote{\url{https://github.com/mkplummer/Imber}}. The code employs Bayesian methods, specifically dynamic nested sampling \citep{Skilling2004,Skilling2006,Higson2019}, to find the best fit for each lightcurve using the Python module \texttt{Dynesty} \citep{Speagle2020}. Decomposing lightcurves into their harmonic components provides the capability to directly infer parameters including amplitude, period, phase, and damping ratio for each harmonic.  

We modify the harmonic model presented in \citet{Plummer2024} by adding a damping term to allow for transient atmospheric dynamics as follows,

\begin{equation}\label{eqn:var_model}
    F(t) = C_0+C_1 t + \sum_i^N A_i \sin (2 \pi t/P_i + \phi_i) \exp{(- \lambda_i t)},
\end{equation}

where $C_O$ is a bias, $C_1$ is the slope that accounts for variations on time scales that exceed the observation period, and $t$ is time in hours. Each model is composed of the sum of $N$ harmonic modes (the order denoted by $i$) for which $A_i$ is amplitude, $P_i$ is period, $\phi_i$ is phase shift, and $\lambda_i$ is decay rate, which is approximated,

\begin{equation}
    \lambda_i \approx 2 \pi \zeta_i / P_i,
\end{equation}

where $\zeta_i$ is the damping ratio for which zero indicates an undamped oscillation and values greater than one result in an overdamped wave.

\subsection{Fitting Scheme}\label{ssec:model:fitting_scheme}

For the Bayesian fit, we use flat uniform priors for our free parameters: amplitude ($0\%<A_i<5\%$), period ($0<P_i<3 \ \rm{h}$), damping ($0<\zeta_i<1$), phase shift ($0^{\circ}<\phi_i<360^{\circ}$), bias ($1.0\pm 0.25$), and slope ($\pm0.1\,\rm{h^{-1}}$). 

To determine the best-fitting model, we compare each inferred light curve's Bayesian logarithmic evidence ($\log \mathcal{Z}$) and Bayesian Information Criterion (BIC). 

Bayesian evidence ($\mathcal{Z}$) derives from the denominator of Bayes' theorem,

\begin{equation}\label{eqn:bayesrule}
P\big(\Theta|D,M\big) = \frac{P\big(D|\Theta,M\big) \ P\big(\Theta|M\big)}{P\big(D|M\big)},
\end{equation}

where $\Theta$ denotes inferred parameters, $D$ represents observed data, and $M$ the applied model. Per Bayesian standard, $P\big(\Theta|D,M\big)$ is the posterior distribution, $P\big(D|\Theta,M\big)$ is the likelihood function, $P\big(\Theta|M\big)$ is the prior distribution, and $P\big(D|M\big) = \mathcal{Z}$ is Bayesian evidence \citep[e.g.,][]{Speagle2020}.

BIC is computed in the traditional manner \citep{Kass&Raftery1995},

\begin{equation}
    \text{BIC} = \chi^2 + m \ln(n) = \sum_{i=1}^{n}\frac{(O_i-M_i)^2}{\sigma^2} + m \ln(n), 
\end{equation}

where $m$ is the number of free parameters, $n$ is the number of observations, $O_i$ and $M_i$ are the $i^{\rm th}$ observation and model data, and $\sigma$ is computed uncertainty. The model with the highest $\log \mathcal{Z}$ and lowest BIC value is considered the best fit.

\subsection{Best Fits}\label{ssec:models:best_fits}

The best-fitting models for both JWST NIRISS/SOSS and NIRSpec/PRISM data are displayed as variability maps (dynamic spectra) in the middle row of Figure~\ref{fig:dynamic_spectra} with residuals between the observation and best-fit models shown in the bottom row of the same figure. The NIRISS/SOSS data's larger residuals ($0.00 \pm0.215 \%$) are likely due to the data's larger variability amplitude and marginally lower binned signal-to-noise ratio. The data appears to approximate white noise but lower amplitude harmonics may be hidden. The residuals in the NIRSpec/PRISM  are smaller in amplitude ($0.00 \pm0.075 \%$) and appear to contain structured harmonics of $k\gtrsim3$. As the Bayesian framework in \texttt{Imber} has difficulty converging on models with $\gtrsim4$ harmonics, this signal is likely due to an unmodeled, high-order harmonic. It should be noted that the signal appears at nearly all wavelengths (and therefore pressure levels), suggesting either a source at deep pressure levels (e.g., iron cloud modulation) or systematics. More details on the results of these fits can be found in \S \ref{ssec:hamonic_analysis}.


\subsection{Contribution Function}\label{ssec:models:CF}

To map the inferred harmonic modes from wavelength-space to pressure-level space in \S \ref{sec:Results}, we use a contribution function. Contribution functions show the emitted radiation at each pressure level in wavelength-space and can be computed in a similar manner as \citet{Lothringer2018},

\begin{equation}\label{eqn:CF}
    C(p,\lambda) = S(p,\lambda) *\frac{d\tau(p,\lambda)}{d\ln p} e^{-\tau(p,\lambda)}
\end{equation}

where $p$ is pressure, $\lambda$ is wavelength, $S(p,\lambda)$ is the source function, and $\tau(p,\lambda)$ is the pressure- and wavelength-dependent optical depth.



For this work, we select a cloudless, $\rm{T_{eff} = 1150~K}$, $\rm{log(g) = 4.5}$ Sonora Bobcat \citep{Marley2021} contribution function (see Figures \ref{fig:NIRISS_Wavelength} and \ref{fig:NIRSPec_Wavelength}). The temperature and surface gravity for the model are selected based on findings of \citet{Gagne2017} and \citet{Vos2023}. For similar reasons as \citet{McCarthy2025}, we make the decision to use a cloudless model to probe the deepest observable pressure levels on SIMP~J0136. This approach is supported by the likelihood that planetary-mass objects are composed of an admixture of cloudless and cloudy regions, based on a study of 2MASS~1207\,b, an object with similar temperature and surface gravity to SIMP~J0136 \citep{Zhang2025}. It should be noted that we do not necessary expect harmonic parameters to increase or decrease in the same manner as the pressure they probe (i.e., the deepest pressure should not necessarily have to have the shortest periods and lowest amplitudes). 

\par As adapting a single contribution function may bias results \citep[e.g., Figure 4 in ][]{Dobbs-Dixon2017}, we compared the vertical variation maps described in \S \ref{ssec:composite_maps} to maps created using the retrieved contribution function in Figure 3 of \citet{Nasedkin2025}. We found that both maps contained the same key features with only minor differences that did not detract from our conclusions.



\section{Results}\label{sec:Results}

\subsection{Harmonic Analysis}\label{ssec:hamonic_analysis}

\par To investigate the rotational modulation and atmospheric structure of SIMP~J0136, we applied 1-D harmonic wave decomposition (see \S \ref{ssec:models:wave_model}) to the lightcurves obtained from both the NIRISS/SOSS and NIRSpec/PRISM instruments. By decomposing the observed variability into sinusoidal components, we retrieved wave periodicity, amplitude, phase shift, and damping ratios as a function of wavelength and associated atmospheric depth. These results allow us to examine how different spectral regions and molecular absorption (and potentially emission) features contribute to the observed spectrophotometric variability of SIMP~J0136. By testing models with varying harmonic complexity, we assess the presence of layered atmospheric dynamics and chemical structures. 

\subsubsection{NIRISS/SOSS Harmonic Results} \label{ssec:results:NIRISS}

\begin{figure*}
\centering
\includegraphics[width=0.95\textwidth]{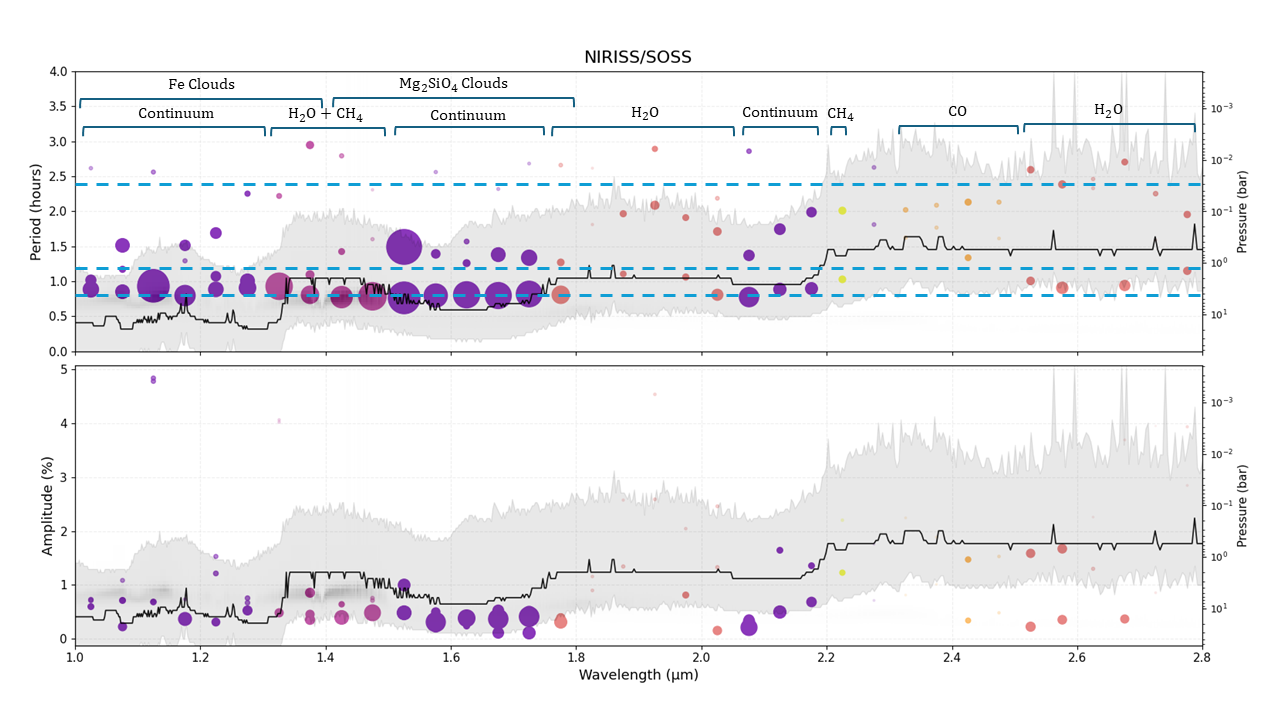}
\caption{\label{fig:NIRISS_Wavelength} NIRISS/SOSS comparison of retrieved harmonic periods and amplitudes as a function of wavelength. Each data point denotes an inferred harmonic mode with either 2 or 3 modes per wavelength (contingent on the best-fit). Data points' sizes are inverse to their relative uncertainties. As seen on the left y-axis: \textbf{Top Row:} Period vs. Wavelength. It can be seen that $k=3$ ($\sim0.8$~h) harmonics correspond to the deepest pressure levels ($\gtrsim1~$bar). \textbf{Bottom Row:} Amplitude vs. Wavelength.  Color denotes the dominant spectral absorption feature in each bin: Continuum (purple), H$_{2}$O + CH$_{4}$ (magenta), H$_{2}$O (red), CO (orange), CH$_{4}$ (yellow). The teal, dashed horizontal lines denote 2.4 h ($k=1$), 1.2 h ($k=2$), and 0.8 h ($k=3$) periodicities. The black line corresponds to the maximum flux from the Sonora Bobcat \citep{Marley2021} contribution function seen at each wavelength and the associated pressures as seen on the right y-axis. The gray shaded region represents $10^{-3}$ of the maximum contribution function flux.}
\end{figure*}

\par Considering NIRISS/SOSS lightcurves that span from 1.0 to 2.8~$\mu$m with 0.05~$\mu$m intervals (36 lightcurves), we identify periodic signals for each bin by fitting harmonic models using the Bayesian framework described in \S \ref{sec:Models}. We assume that variations in specific wavelength intervals corresponding to known molecular bands reflect variations in the absorption of those species. If the observed molecular variability was only due to the presence of clouds muting spectral features, all species would be uniformly affected. However, we see in \S \ref{ssec:unique_maps} that the presence of clouds leads to enhanced absorption for methane and reduced absorption for H$_2$O and CO. This allows us to connect the retrieved rotational periodicities, amplitudes, and phase shifts with changes in atmospheric composition and depth. 



\par Our harmonic analysis reveals best-fit harmonic structure varies by wavelength. Short wavelengths (1.0 to 1.45~$\mu$m) are predominantly described by three-harmonic models with mixed damping, the mid-range (1.5 to 2.1~$\mu$m) shows no strong preference between two- and three-harmonic fits, and longer wavelengths (2.15 to 2.8~$\mu$m) are consistently best fit by two-harmonic models without a systematic damping trend. Figure~\ref{fig:NIRISS_Wavelength} presents retrieved periods (Top Row) and amplitudes (Bottom Row) across the NIRISS/SOSS spectral range. Points are color-coded on the basis of the dominant molecular absorption feature in each bin and inversely sized on the basis of retrieved posterior uncertainty.

\par The lowest-order periodicity across NIRISS/SOSS wavelengths is the $k=1$ harmonic at $\sim$2.4~h, consistent with SIMP~J0136's rotation \citep{Yang2016}. These harmonics tend to have the highest uncertainty as the period is of similar order as the observational period.


CO bands, and to a lesser extent H$_2$O, show retrieved periods that consistently fall between the fundamental rotation ($k=1$, 2.4 h) and its first harmonic ($k=2$, 1.2 h), producing beating patterns with characteristic periodicities spanning 1.2 to 2.4 h at pressure levels ranging from $\sim1$ to $10^{-2}~$bar. These beating patterns were also seen in \citet{Apai2017, apai21, Fuda:2024} for WISE~1049\,B and \citet{Zhou2022} for VHS$~$1256\,b.

\par Higher-order ($k>2$) variability traces deeper atmospheric layers as seen in Figure~\ref{fig:NIRISS_Wavelength}. The pressure levels associated with the maximum flux from the spectral contribution function (see \S \ref{ssec:models:CF}) are overlaid onto the retrieved components. From 1.0 to 1.8 $\mu$m and $\sim 2.0$ to 2.2~$\mu$m, the $k=3$ (0.8 h) harmonic is preferentially retrieved at pressure levels $\sim 10$ bar, consistent with variability originating from deeper atmospheric layers. It is noteworthy that the $k=3$ harmonics also have the smallest posterior uncertainty.

\par Amplitudes are generally low across 1 to 1.8 $\mu$m ($<1.5\%$) with isolated outliers at 1.15 and 1.35 $\mu$m. However, amplitudes increase with the onset of H$_2$O absorption near 1.8 $\mu$m, with both variability and uncertainties growing toward longer wavelengths. The H$_2$O absorption region from 2.5 to 2.8~$\mu$m reaches the highest amplitudes for the NIRISS/SOSS range.






\subsubsection{NIRSpec/PRISM Harmonic Results} \label{ssec:Results:NIRSPEC}

\begin{figure*}
\centering
\includegraphics[width=1\textwidth]{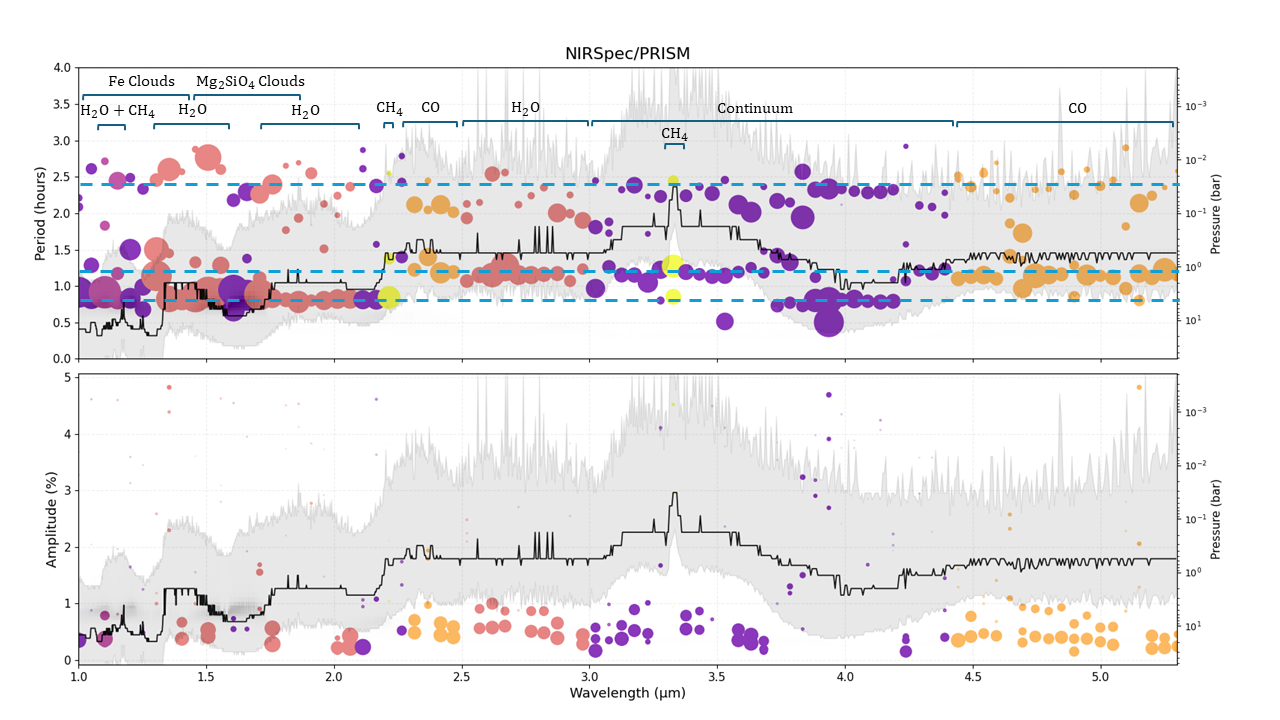}
\caption{\label{fig:NIRSPec_Wavelength}
NIRSpec/PRISM comparison of retrieved harmonic periods and amplitudes as a function of wavelength. Each data point denotes an inferred harmonic mode with either 2 or 3 modes per wavelength (contingent on the best-fit). Data points' sizes are inverse to their relative uncertainties. As seen on the left y-axis: \textbf{Top Row:} Period vs. Wavelength. It can be seen that $k=2$  ($\sim1.2$~h) harmonics correspond to $\sim1~$bar while $k=3$ ($\sim0.8$~h) harmonics correspond to the deepest pressure levels ($\gtrsim1~$bar).  \textbf{Bottom Row:} Amplitude vs. Wavelength. Color denotes the dominant spectral absorption feature in each bin: Continuum (purple), H$_{2}$O + CH$_{4}$ (magenta), H$_{2}$O (red), CO (orange), CH$_{4}$ (yellow). The teal, dashed horizontal lines denote 2.4 h ($k=1$), 1.2 h ($k=2$), and 0.8 h ($k=3$) periodicities. The black line corresponds to the maximum flux from the Sonora Bobcat \citep{Marley2021} contribution function seen at each wavelength and the associated pressures as seen on the right y-axis. The gray shaded region represents $10^{-3}$ of the maximum contribution function flux.}
\end{figure*}

\par Considering a second epoch of JWST observation and extending our harmonic analysis across a wider wavelength range, we applied the same periodic modeling procedure to the lightcurves from NIRSpec/PRISM, covering 1.0 to 5.3~$\mu$m in 0.05~$\mu$m bins (86 lightcurves). 






Each NIRSpec/PRISM binned lightcurve was tested to find the best-fit model (see Figure~\ref{fig:dynamic_spectra}, Middle Row). Figure~\ref{fig:NIRSPec_Wavelength} presents the retrieved periods (Top Row) and  amplitudes (Bottom Row) across the NIRSpec/PRISM range. Wavelengths from 1.0 to 1.5~$\mu$m were predominantly best fit by three-harmonic models. Between 1.5 to 2.1~$\mu$m, there was no strong preference between two- and three-harmonic fits, nor any consistent trend in damping configuration. For wavelengths beyond 2.1~$\mu$m, two-harmonic models became increasingly dominant. 


Beating patterns with inferred periods between the $k=1$ (2.4 h) and $k=2$ (1.2 h) harmonics appear to be dominant from 2.25 to 3.7~$\mu$m, encompassing CO, H$_{2}$O and continuum lines, and ranging in pressure from 1 to $10^{-3}$ bar. The clearest beating patterns appear in the CO lines and the H$_2$O band between 2.5 and 3.0$~\mu$m.



\par The retrieved $k=3$ harmonic is again associated with deeper layers at $\sim 10$ bar, consistent with NIRISS/SOSS results that locate higher-order variability at greater pressures. At pressures $\lesssim1~$bar at 2.25 $\mu$m, models with 2 harmonics begin to dominate. Comparing the inferred amplitudes, there is increased variability in the $k=1$ and $k=2$ harmonics, especially in the deeper regions from 1 to 2.25 $\mu$m and the transition region from 3.5 to 4.4 $\mu$m. It should be noted that the CH$_4$ absorption bin at 3.3 $\mu$m is associated with the highest altitude (lowest pressure) observed from this data.

\par Mirroring the NIRISS/SOSS results, the H$_2$O absorption bands have large amplitudes; however, both the wavelengths associated with cloud modulation (1.0 to 1.8~$\mu$m) and the high-altitude continuum ($\sim3.6$ to 4.0~$\mu$m) also possess large amplitude waves, contrasting with the NIRISS data. The 1.8 to 3.45~$\mu$m range remains suppressed in amplitude while the CO band from 4.3 to 5.3~$\mu$m shows the smallest amplitudes.


\subsection{Vertical Variability Mapping via Harmonic Waves}

\par Time-resolved, multi-spectral data such as those collected by JWST NIRISS and NIRSpec instruments provide the opportunity to map the dynamic atmospheric physics of brown dwarfs and exoplanets at varying pressure levels. Using a method first demonstrated in \citet{Akhmetshyn2025}, we use the best-fit harmonic models (Figure~\ref{fig:dynamic_spectra}, Middle Row), along with a model contribution function (see \S \ref{ssec:models:CF}), to create vertical flux variability heat maps (pressure vs. time) of the atmosphere of SIMP~J0136. These vertical variability maps demonstrate the flux variability at each atmospheric layer, the interaction between layers, and unique spectral contributions to the object's dynamic atmosphere.

\subsubsection{Composite Vertical Variability Maps}\label{ssec:composite_maps}

\par To build composite vertical variability maps, we start by considering a time-varying contribution function of similar form as Equation \ref{eqn:CF},

\begin{equation}\label{eqn:CF_with_time}
    C(p,\lambda,t) = S(p,\lambda,t) *\frac{d\tau(p,\lambda,t)}{d\ln p} e^{-\tau(p,\lambda,t)}.
\end{equation}

Adopting a similar approach as that seen in phase-curve mapping literature \citep[e.g.,][]{Irwin2020,Al-Refaie2021,Changeat2021,Changeat2022,Chubb2022}, we assume that opacity ($\kappa$) is time-independent and that temporal or longitudinal variability is only expressed through the source function (a necessary assumption so that we can employ forward, grid-based, and self-consistent atmospheric models) such that

\begin{equation}
    \kappa(p,\lambda,t) = \kappa(p,\lambda)
\end{equation}

and therefore,

\begin{equation}
    \tau(p,\lambda,t) = \tau(p,\lambda),
\end{equation}

resulting in a pressure- and wavelength-dependent transmission term,

\begin{equation}
    T (p,\lambda) \equiv \frac{d\tau(p,\lambda)}{d\ln p} e^{-\tau(p,\lambda)}.
\end{equation}

We can then restate Equation \ref{eqn:CF_with_time} as follows,

\begin{equation}\label{eqn:CF_time_with_flux}
    C(p,\lambda,t) =S(p,\lambda,t)~T(p,\lambda).
\end{equation}

Again borrowing from phase-curve mapping literature, we approximate the source function as a Planck function, $B_{\lambda}[T]$, and linearize about a reference temperature, $T_0$ \citep[see e.g.,][]{Mihalas1984,Goody1989}. We then parameterize the components of the temperature perturbation, $\delta T$, as separable variables \citep[see e.g., ][]{Rodgers2000}. Linearization requires that $|\delta T| \ll T_0$, a requirement met by the small \% variability in our data. The result of this parameterization is that the source function is correspondingly factorized as follows:

\begin{equation}\label{eqn:source_fun}
    S(p,\lambda,t) \approx S_0(p,\lambda) ~s(\lambda,t) \approx S_0(p,\lambda) ~F(\lambda,t) ,
\end{equation}

where $S_0(p,\lambda)$ represents a component of the mean atmospheric temperature structure and $s(\lambda,t)$, the spectral variability of the emitted radiation. The spectral flux, $s(\lambda,t)$, is modeled using $F(\lambda,t)$, the normalized, best-fit spectral lightcurves (see Equation \ref{eqn:var_model}). 

Combining Equations \ref{eqn:CF_time_with_flux} and \ref{eqn:source_fun} results in a time-varying contribution function in which the time-dependence is isolated to the spectral flux, $F(\lambda,t)$,

\begin{equation}
    C(p,\lambda,t) = S_0(p,\lambda)~T(p,\lambda)~F(\lambda,t),
\end{equation}

and it can be seen that $S_0(p,\lambda)~T(p,\lambda)$ is the time-independent (mean) contribution function, $C(p,\lambda)$.

\begin{figure*}
\centering
\includegraphics[width=1.0\textwidth]{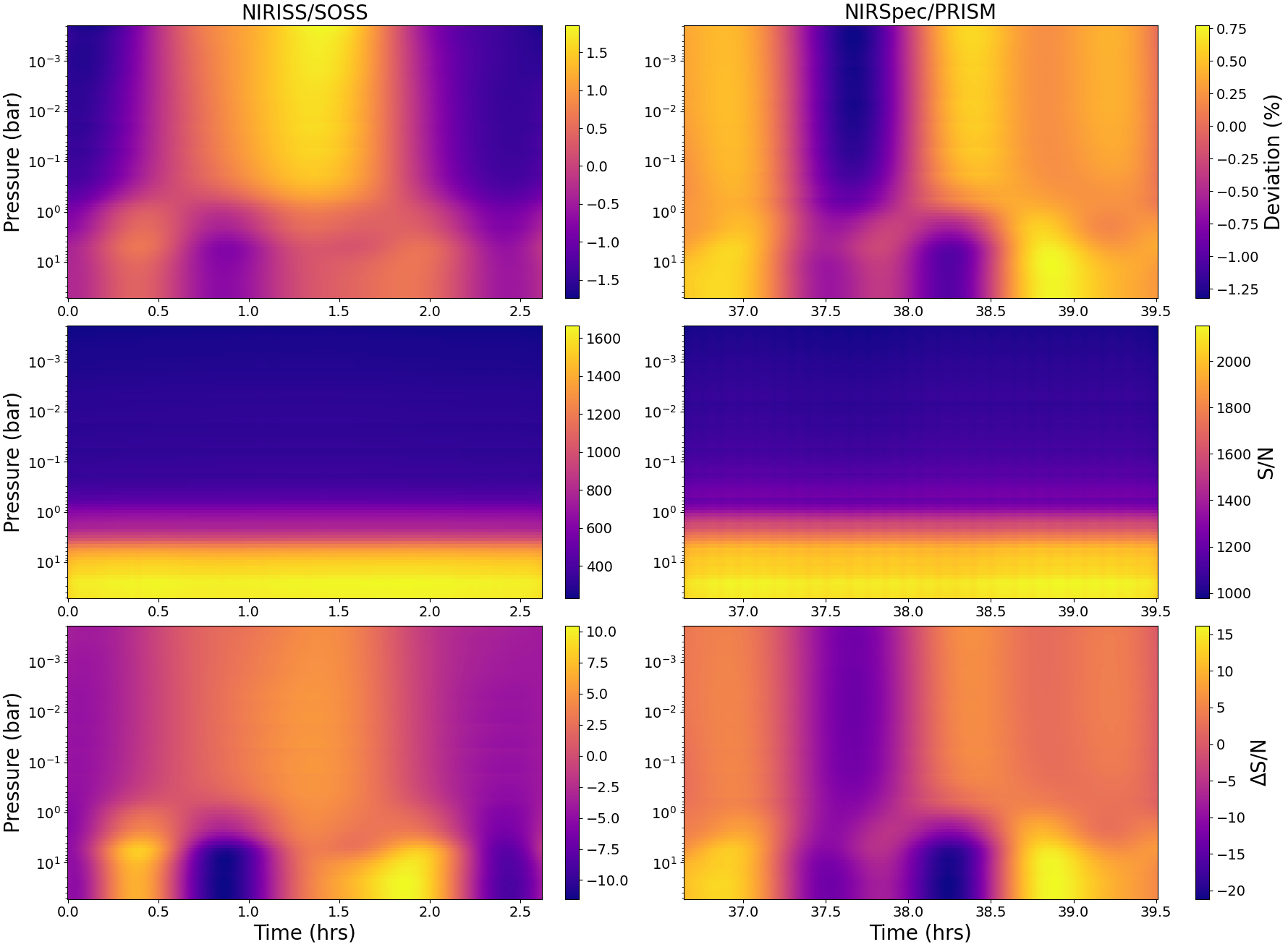}
\caption{\label{fig:Composite_Maps} NIRISS/SOSS (left) and NIRSpec/PRISM (right) vertical variability maps for SIMP~J0136. (Top Row) Composite maps including variation in \% deviation across all observed wavelengths. (Middle Row) S/N vertical map. Pressure levels $\gtrsim 1~$bar have the highest S/N corresponding to higher observational S/N in wavelengths $\lesssim 1.8~\mu$m (Bottom Row) S/N-weighted composite variability maps. Stratification into $\gtrsim2$ interacting layers can be observed. As rotational period is $\sim 2.4$ h, longitudinal wrapping can be observed within each observation. Visually, atmospheric evolution can be seen over the $\sim 35~$h between observations. 
}
\end{figure*}

With these results in mind, we compute the pressure- and time-dependent variability, $V(p_i,t)$, at each time-step and pressure level ($p_i$) and numerically sum the scaled spectral contribution across $N_{\lambda}$ wavelength bins,

\begin{equation} \label{eqn:vert_var}
    V(p_i,t) = \sum^{N_{\lambda}}_j C(p_i,\lambda) F(\lambda,t) \Delta \lambda_j,
\end{equation}

resulting in a 1-D flux array in pressure-level space. The computation is then repeated at each time-step, followed by normalization of flux at each pressure level, resulting in a composite vertical variability map (Figure~\ref{fig:Composite_Maps}, Top Row) in terms of \% deviation. The composite $\%$ deviation maps broadly match the dynamic spectra (variability maps) in Figure \ref{fig:dynamic_spectra}, and the vertical variability maps can be seen as a (model-weighted) coordinate transformation of the flux heat map from time vs. wavelength to time vs. pressure.


\par To quantify the significance of the observed features in the retrieved maps, we scale the composite variability map (Figure \ref{fig:Composite_Maps}, Top Row) by the S/N in pressure vs. time space (Figure \ref{fig:Composite_Maps}, Top Row). The S/N map is computed in a similar manner as the composite map, using Equation \ref{eqn:vert_var}, but the normalized lightcurve flux is replaced with the S/N for each spectral bin and time-step. Comparing the S/N values in Figure \ref{fig:Composite_Maps}, Bottom Row to the binned values in Figure \ref{fig:spectra}, Bottom Row, shows agreement for both NIRISS/SOSS and NIRSpec/PRISM datasets. The resulting S/N-weighted map (Figure \ref{fig:Composite_Maps}, Bottom Row) shows the highest deviation ($\rm{\Delta S/N}$) in the deeper pressure levels ($\gtrsim 1~$bar) due to the high S/N in these regions. The shallow layer ($\lesssim 1~$bar) exhibits smaller $\rm{\Delta S/N}$ due to its lower S/N despite possessing larger $\%$ deviation.

\par SIMP~J0136's composite vertical variability maps of the NIRISS/SOSS and NIRSpec/PRISM data reveal both similarities in the atmospheric structure observed at each epoch and apparent qualitative differences between the two observations (separated by $\sim 35~$h or $\sim15$ rotations). Both NIRISS/SOSS and NIRSpec/PRISM maps show two dynamic interacting atmospheric layers with time-varying boundaries: (1) a lower pressure (higher altitude) overlying layer comprised of lower order harmonics and (2) a deeper layer ($\gtrsim 1~$bar) corresponding to expected cloud formation regions of forsterite and potentially iron \citep[see e.g.,][]{Vos2023,McCarthy:2024,McCarthy2025}. The time-varying pressure level boundary between the deep and overlying atmospheric layers suggests variations in vertical cloud structure.

However, visual differences between the two observations' maps can be seen, suggesting complex weather patterns evolving over a relatively short timescale. In the NIRISS/SOSS observation, the deep layer's harmonic pattern is alternating single peaks of minima-maxima-minima followed by a double-peaked maximum. For the NIRSpec/PRISM observation, there are two pairs of double-peaked minima and maxima. The overlying layer evolves in complexity with the NIRSpec/PRISM observation, including a double-peaked maxima not observable in the NIRISS/SOSS data. Note that the NIRSpec/PRISM data extend to longer wavelengths and, therefore, lower pressure levels than the NIRISS/SOSS. As a result, the overlying layer in the NIRSpec/PRISM observation possesses more retrieved structure.



\subsubsection{Spectral Features' Deviations from Bulk Variability}\label{ssec:unique_maps}

\par The broad NIR spectral coverage of JWST NIRISS/SOSS and NIRSpec/PRISM allows us to map spectral features' deviations from bulk (or composite) variability and investigate that deviation's ties to distinct atmospheric phenomena. For example, we can determine how wavelengths corresponding to iron ($1.0$ to $1.4~\mu$m) and forsterite ($1.4$ to $1.8~\mu$m) cloud modulation \citep{Vos2023,McCarthy2025} or known molecular absorption bands (H$_2$O, CO, CH$_4$) deviate from bulk variability over the period of observation.

\begin{figure*}
\centering
\includegraphics[width=0.9\textwidth]{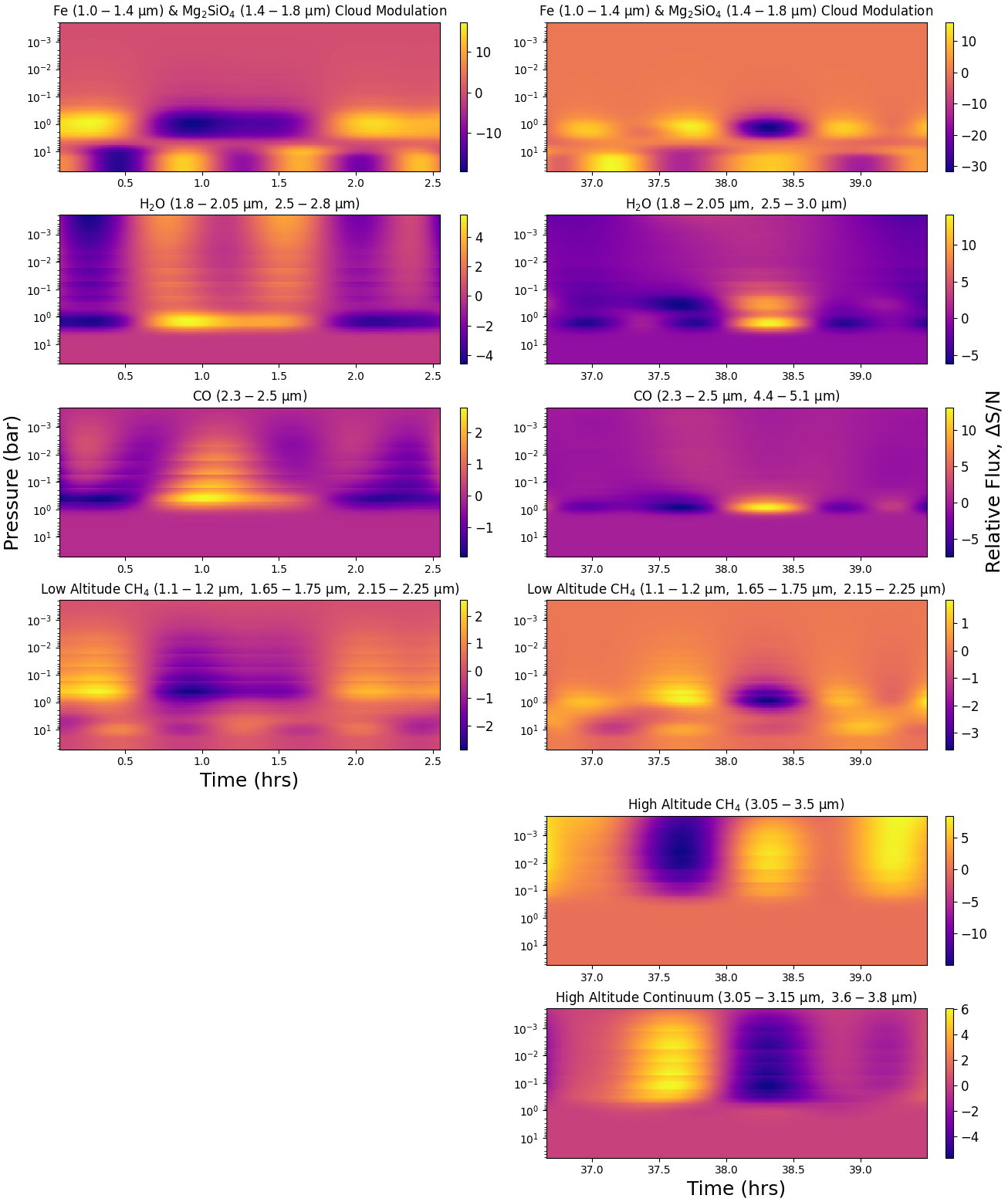}
\caption{\label{fig:Vertical_Var_Map} NIRISS/SOSS (left) and NIRSpec/PRISM (right) spectral features' deviations from bulk variability for SIMP~J0136. \textbf{Top Row:} Presumed iron (lower layer, $1.0$ to $1.4~\mu$m) and forsterite (upper layer, $1.4$ to $1.8~\mu$m) cloud modulation. Wavelengths based on models and previous observations of cloud-influenced pressure levels and wavelengths for SIMP~J0136 \citep[see e.g.,][]{Vos2023,McCarthy2025}. \textbf{Second Row:} H$_2$O deviation map; anti-correlated with forsterite cloud modulation. \textbf{Third Row:} CO deviation map; anti-correlated with forsterite cloud modulation. \textbf{Fourth Row:} Low-altitude CH$_4$ deviation map; correlated with forsterite cloud modulation. \textbf{Fifth Row:} (NIRSpec/PRISM only) High-altitude CH$_4$ deviation map. Deviation possesses similar harmonic periodicity of low-altitude CH$_4$ feature above cloud tops but with reverse sign, tentatively indicating emission at higher altitudes. \textbf{Bottom Row:} (NIRSpec/PRISM only) High-altitude continuum deviation; correlated with forsterite cloud modulation.
}
\end{figure*}

Maps depicting specific spectral regions' deviations from bulk variability can be seen in Figure~\ref{fig:Vertical_Var_Map}. These maps are created by starting with the composite map (Figure~\ref{fig:Composite_Maps}) and then subtracting from it a second map comprising the wavelengths available for which we are not interested for a particular test, resulting in a spectrally-unique, \% deviation map. This \% deviation map is then scaled by the average S/N for the wavelengths of interest. To maximize the S/N, an additional round of harmonic fitting was performed with 8.71 min temporal binning (for both observations) and spectral binning based on the entire wavelength region of interest (e.g., 1.4 to 1.8$~\mu$m for forsterite clouds). 

It is important to note that maps showing deviations of individual spectral features from the bulk variability do not necessarily indicate genuine, feature-specific variability. They reflect differences in the flux relative to the composite (bulk) signal. For example, a molecular absorption band may exhibit no intrinsic time-dependent behavior, but if bulk variability is subtracted from the associated dynamic spectra, the resulting residuals would exhibit an induced anti-correlation. Despite this, deviations from the bulk variability can be meaningful and provide insight into wavelengths with distinct time-varying behavior.

The deviation maps for wavelengths expected to correspond to mineral cloud modulation \citep[i.e., iron and forsterite,][]{Vos2023} demonstrate clear harmonic patterns for both the NIRISS/SOSS and NIRSpec/PRISM datasets (see Figure~\ref{fig:Vertical_Var_Map}, Top Row). The lower layer ($\gtrsim3~$bar) is suspected to be iron cloud modulation with maximum signal-to-noise (S/N$_{\rm{max}}$) deviations of 18.5 for NIRISS/SOSS and 16.1 for NIRSpec/PRISM observations. The upper forsterite layer (centered at \hbox{$\sim$1\,bar}) is approximately anti-correlated with the underlying iron layer and has S/N$_{\rm{max}}$ deviations of 16.3 and 26.5 for NIRISS/SOSS and NIRSpec/PRISM respectively.  

\par For the forsterite variation, the NIRISS/SOSS data include a double-peaked dark patch but the NIRSpec/PRISM observations contain only a single dark patch. The iron cloud regions do not appear to possess double-peaked features. For these maps, the dark and bright patches are presumed to be longitudes of enhanced or reduced cloud coverage, respectively, and will be discussed further in \S \ref{ssec:discussion:anticorr}. 

Deviation with S/N$_{\rm{max}}$ = 5.5 (NIRISS/SOSS) and 14.0 (NIRSpec/PRISM) in wavelengths corresponding to dominant H$_{2}$O absorption bands ($1.8$ to $2.05~\mu$m, $2.5$ to $2.8~\mu$m for NIRISS/SOSS and $2.5$ to $3.0~\mu$m for NIRSpec/PRISM) is anti-correlated with the suspected $1.4$ to $1.8~\mu$m forsterite cloud modulation for both data sets (see Figure~\ref{fig:Vertical_Var_Map}, Second Row). H$_{2}$O dark spots (enhanced absorption regions) anti-correlate with the $1.4$ to $1.8~\mu$m dark spots while H$_{2}$O bright spots (reduced absorption regions) correlate with the $1.4$ to $1.8~\mu$m dark spots. H$_{2}$O deviations are strongest (highest S/N) at the $\sim1~$bar pressure level, and are constrained to $\gtrsim 100~$mbar in the NIRSpec/PRISM data, but are relatively unconstrained in terms of a lower pressure bounds for NIRISS/SOSS data due to limited wavelength coverage.

Similar to H$_{2}$O, deviations in the CO absorption bands ($2.3$ to $2.5~\mu$m, with the NIRSpec/PRISM map also including the $4.4$ to $5.1~\mu$m CO band) are anti-correlated with the $1.4$ to $1.8~\mu$m modulation and show the strongest deviations at the $\sim1~$bar pressure level (Figure~\ref{fig:Vertical_Var_Map}, Third Row). Evidence for deviation is marginal (S/N$_{\rm{max}}$ = 2.78) for the NIRISS/SOSS data but strong (S/N$_{\rm{max}}$ = 14.0) for NIRSpec/PRISM observations. The NIRISS/SOSS map extends up to the $1~$mbar level with a wedge-like pattern. The NIRSpec/PRISM map, on the other hand, is constrained to $\sim1~$bar pressure levels. This difference is again likely due to the constraining effect of the wider spectral coverage of NIRSpec/PRISM.   

As opposed to the H$_{2}$O and CO maps, the wavelengths corresponding to lower-altitude (deeper pressure levels) CH$_{4}$ absorption bands ($1.1$ to $1.2~\mu$m, $1.65$ to $1.75~\mu$m, $2.15$ to $2.25~\mu$m) are correlated with $1.4$ to $1.8~\mu$m modulation in both observations (see Figure~\ref{fig:Vertical_Var_Map}, Fourth Row). Evidence for lower-altitude CH$_4$ deviation is weaker than for H$_{2}$O and CO, with marginal evidence (S/N$_{\rm{max}}$ = 2.9) for NIRISS/SOSS data and tentative evidence (S/N$_{\rm{max}}$ = 3.6) for the NIRSpec/PRISM maps. As with the CO maps, the NIRISS/SOSS deviations extend up to the $1~$mbar level while the NIRSpec/PRISM deviations are constrained to the $\sim1~$bar pressure level.

As the NIRSpec/PRISM data extends to 5.3$~\mu$m, it provides access to higher-altitude spectral features (as shown in the contribution function in Figures \ref{fig:NIRISS_Wavelength} and \ref{fig:NIRSPec_Wavelength}). The wavelengths from $3.05 \ \rm{to} \ 3.5~\mu$m bracket a CH$_{4}$ feature reaching $\lesssim1~$mbar pressure levels. The high-altitude CH$_{4}$ map (see Figure~\ref{fig:Vertical_Var_Map}, Fifth Row) has a distinct (and potentially anti-correlated) deviation pattern (S/N$_{\rm{max}}$ = 14.9) compared to the lower-altitude CH$_{4}$ absorption bands. Longer wavelengths also include the relatively high-altitude spectral continuum (3.05 to 3.15 $\mu$m, 3.6 to 3.8 $\mu$m, see Figure~\ref{fig:Vertical_Var_Map}, Bottom Row) with deviations (S/N$_{\rm{max}}$ = 6.0) appearing to be correlated with the $1.4$ to $1.8~\mu$m modulation and anti-correlated with high-altitude CH$_4$ deviations.

\section{Discussion}\label{sec:Discussion}

\subsection{Odd Harmonics in the Forsterite Cloud Layer}\label{ssec:disc:harmonics}

Odd ($k=3$) harmonics in SIMP~J0136's spectrophotometry (see Figures \ref{fig:NIRISS_Wavelength} and \ref{fig:NIRSPec_Wavelength}, Top Row) are preferentially found at wavelengths corresponding to the deepest probed pressure layers as well as iron and forsterite cloud modulation (1.0 to 1.8$~\mu$m). In the NIRSpec/PRISM data set, the wavelengths expected to be most affected by cloud modulation (1.0 to 1.8$~\mu$m) and the 3.75 to 4.2~$\mu$m continuum also show high variability, suggesting the longer-wavelength continuum may be driven by the underlying cloud layer.  

As initially reported by \citet{Akhmetshyn2025}, odd ($k\geq3$) harmonics in the NIRISS/SOSS data may indicate North/South asymmetry in the surface brightness map of SIMP~J0136 \citep[depending on the limb darkening law;][]{luger21d}. In the absence of limb darkening, even spherical-harmonic degrees ($l$) paired with odd azimuthal orders $m$ (which correspond to the 
longitudinal wavenumber $k$) signify a North–South asymmetry in the surface-brightness distribution \citep{Cowan&Fuentes2013}, an assumption adopted by \citet{Akhmetshyn2025}. The situation changes, however, once limb darkening is included: linear and quadratic limb–darkening laws effectively increase the contributing spherical-harmonic degree $l$ by one and two, respectively \citep{luger21d}. Consequently, a linear limb–darkening law can generate apparent odd-$m$ (odd-$k$) signals even for a North–South symmetric map, whereas under a quadratic law—or in the limit of negligible limb darkening—the presence of such odd 
modes once again implies genuine hemispheric asymmetry. Additional constraints on the limb–darkening characteristics of SIMP~J0136 and similar objects will therefore be essential for robust interpretation of their light-curve morphologies.


If the odd harmonics are indeed due to North/South asymmetry, then they also support an inclined viewing geometry \citep{Cowan&Fuentes2013}, consistent with the estimate of $80^{\circ} \pm 10^{\circ}$ \citep{Vos2017}. Furthermore, this asymmetry could help lift hemispheric ambiguity in future Doppler imaging efforts with ELT-class telescopes with high-resolution spectrographs \citep[see e.g.,][]{Plummer&Wang2023}, making SIMP~J0136 an ever more compelling target for such work.

Because brown dwarfs have been found to be redder, and therefore cloudier, at equatorial latitudes than polar ones \citep{Vos2017,Suarez&Metchev2022,Suarez2023}, the odd harmonics at the deepest pressure levels (corresponding to the cloud formation region) potentially suggest the North/South asymmetry may lie in the mineral cloud modulation region and be concentrated near the equator. Multi-rotational NIR spectroscopic observations of SIMP~J0136 would help to constrain higher order odd harmonics, decrease the uncertainty in retrieved harmonic parameters, and elucidate the dynamical structure of L/T transition objects.

\subsection{Cloud-Driven Temperature and Chemical Structure}\label{ssec:discussion:anticorr}

The anti-correlation between the 1.4 to 1.8$~\mu$m forsterite cloud modulation region and H$_{2}$O and CO absorption bands, as well as the correlation of CH$_{4}$ with the same cloud region (see Figure~\ref{fig:Vertical_Var_Map}), suggests that forsterite clouds influence the atmospheric chemical structure of SIMP~J0136, potentially by driving variations in temperature structure and chemical mixing ratios. H$_{2}$O and CO$_{2}$ display reduced absorption at and above the cloud formation level while CH$_4$ exhibits enhanced absorption. The differing behavior of CH$_4$ implies that the explanation for the anti-correlation of H$_{2}$O and CO$_{2}$ deviations with the cloud layers cannot be completely explained due to clouds muting the observability of these absorption features.

We suggest two potential causes for the relationship between the deviations in the cloud modulation region and molecular absorption bands: (1) Cloud modulation influences the temperature structure, and therefore, the chemical conditions of the atmosphere and (2) Cloud formation and dissipation controls the oxygen abundances at co-altitude and overlying pressure levels. Both mechanisms could play a part in determining the dynamic and chemical structure of the upper atmosphere of SIMP~J0136.

Forsterite cloud modulation can be seen to drive the upper atmosphere temperature structure as shown by the correlation between the dark cloud feature in Figure~\ref{fig:Vertical_Var_Map} (Top Row, Right Column) and the high-altitude continuum in Figure~\ref{fig:Vertical_Var_Map} (Bottom Row, Right Column). Regions above the clouds are relatively cooler while gaps in the cloud layer allow heat to escape initially in vertical columns \citep[see e.g.,][]{Showman&Tan2019,tan21a,tan21b}. As CO is the preferred carbon reservoir at higher temperatures (and contrastingly CH$_{4}$ is the preferred reservoir at lower temperatures), cooler regions should exhibit depleted CO and enhanced CH$_{4}$ absorption, matching our findings (Figure~\ref{fig:Vertical_Var_Map}, Third and Fourth Rows) and contributing to carbon disequilibrium chemistry in SIMP~J0136.

To understand the H$_{2}$O deviations in this context, we should consider the following equation \citep[e.g.,][]{Visscher2011},

\begin{equation}\label{eqn:carbon_chemistry}
    \rm{CO+3H_{2} \leftrightarrow CH_{4} + H_{2}O}.
\end{equation}

Assuming an equilibrium condition, as CO is removed from the atmosphere, the balance of the equation shifts to the left, depleting H$_{2}$O abundances.

The sequestration of oxygen in the forsterite cloud formation process may also factor into the observed anti-correlation of the cloud layer and CO/H$_{2}$O. The chemistry governing the forsterite cloud formation process can be understood by the net thermochemical reaction \citep[e.g.,][]{Visscher2010},
\begin{equation}\label{eqn:forsterite_formation}
    \rm{2Mg + 3H_{2}O +SiO = Mg_{2}SiO_{4}(s,l)+3H_{2}},
\end{equation}
where $s$ and $l$ respectively denote a solid or liquid phase. The process consumes H$_{2}$O to form forsterite, potentially explaining the anti-correlation seen in Figure~\ref{fig:Vertical_Var_Map} (Second Row). Depleted oxygen abundances at and above the cloud layer also favor CH$_{4}$ over CO as the preferred carbon reservoir (see Equation \ref{eqn:carbon_chemistry}), again explaining the observed carbon disequilibrium chemistry in SIMP~J0136 as well as the chemical deviation patterns in Figure~\ref{fig:Vertical_Var_Map} (Third and Fourth Rows).

\subsection{High-Altitude Heating Source}\label{ssec:Discussion:auroralheating}

Mapping the deviations in the wavelengths (3.05 to 3.5 $\mu$m) bracketing the high-altitude, $\nu_{3}$ rotational-vibrational CH$_4$ band ($\sim 3.4~\mu$m, see Figure~\ref{fig:Vertical_Var_Map}, Fifth Row) results in harmonics distinct from and largely anti-correlated with the low-altitude CH$_{4}$ bands (see Figure~\ref{fig:Vertical_Var_Map}, Fourth Row). For example, the CH$_4$ absorption feature between 38.0 and 38.5 hr on the low-altitude CH$_4$ map is replaced by a bright spot, which we tentatively interpret as an emission feature in the upper atmosphere, potentially driven by auroral heating from electron precipitation \citep[e.g.,][]{Sandel1982,Melin2011,Muller-Wodarg2012,Pineda2024,McCarthy2025,Nasedkin2025}. In this case, the CH$_4$ abundance appears variable and is influenced and in-phase with underlying cloud modulation, but now we detect the CH$_4$ in emission rather than absorption. 

Distinct, high-altitude CH$_4$ emission also aligns with previous findings for SIMP~J0136 based on the same NIRSpec/PRISM dataset explored in this work. \hbox{K-means} clustering identified unique lightcurve morphology near the CH$_4$ $\nu_3$-band \citep{McCarthy2025}, and a 265\,K temperature inversion was inferred at pressures ($\sim10^{-3}~$bar) corresponding to the same wavelengths \citep{Nasedkin2025}. The necessary power to create such a strong temperature inversion from electron precipitation-driven auroral heating \citep[which is similar in strength to the 300\,K heating found for WISE J1935,][]{Faherty2024}, was computed to be $\sim10^{19}~$W by \citet{Nasedkin2025} while Jupiter-like auroral processes could only produce auroral power on the scale of $10^{17}~$W. 

As discussed in \citet{Nasedkin2025}, other upper atmospheric heating processes such as the breaking of gravity waves \citep{Freytag2010} and Joule heating associated with atmospheric electric currents have been proposed for brown dwarfs based on observations of these mechanisms in outer solar system planets \citep{Cowley2004,Ingersoll2021,Brown2022}. Further studies will be necessary to break the degeneracies between these potential heating sources. 

\section{Summary}\label{sec:Summary}

To understand how mineral cloud modulation, disequilibrium carbon chemistry, and upper atmospheric heating (potentially aurora-driven) each contribute to the well-established NIR variability of the isolated L/T transition, planetary-mass object SIMP~J0136, we analyzed time-resolved spectroscopy gathered from JWST NIRISS and NIRSpec instruments, collected $\sim35~$h apart, and originally presented in \citet{Akhmetshyn2025} and \citet{McCarthy2025}. Performing Bayesian, Fourier fitting via \texttt{Imber}, a Python code developed in \citet{Plummer&Wang2022,Plummer&Wang2023,Plummer2024}, we found the dominant harmonic modes (and associated parameters) for each 0.05$~\mu$m bin. Using these inferred harmonic modes and Sonora Bobcat \citep{Marley2021} atmospheric models, we computed vertical variability maps demonstrating the flux variability at each pressure level and deviations from bulk variability for molecular absorption features (i.e., H$_{2}$O, CO, and CH$_{4}$). We summarize our findings:

\begin{enumerate}

    \item SIMP~J0136's spectral variability appears to evolve over a period of $\sim 35~$h (see \S \ref{ssec:results:NIRISS}, \ref{ssec:Results:NIRSPEC} and Figures \ref{fig:NIRISS_Wavelength} and \ref{fig:NIRSPec_Wavelength}). NIRISS/SOSS spectra show the highest variability ($2.5-5\%$) in H$_{2}$O absorption bands (1.75 to 2.05 $\mu$m and 2.5 to 2.8 $\mu$m) with muted variability at other wavelengths. However, NIRSpec/PRISM spectra have the highest variability at wavelengths $<2~\mu$m (corresponding to wavelengths expected to be affected by iron and forsterite cloud modulation and including the H$_{2}$O absorption band) and also the longer-wavelength continuum (3.75 to 4.2~$\mu$m) corresponding to deep pressure levels. We interpret this variation as dynamic modulation of the forsterite cloud-forming region.
  
    \item Harmonics with periodicities of $\lesssim0.8$h ($k\gtrsim3$) are inferred preferentially at pressure levels corresponding to $\gtrsim1~$bar (the suspected iron and forsterite cloud modulation region, see \S \ref{ssec:results:NIRISS}, \S \ref{ssec:Results:NIRSPEC}, \S \ref{ssec:disc:harmonics} and Figures \ref{fig:NIRISS_Wavelength} and \ref{fig:NIRSPec_Wavelength}). As odd ($k\geq3$) harmonics are potentially indicative of North/South surface map asymmetry (depending on the limb darkening law) and equatorial latitudes are known to be cloudier in brown dwarfs, the observed $k=3$ harmonics may suggest such an asymmetry is primarily contained to the cloud-forming (and likely near-equatorial) regions. 


    \item We detect deviations from the bulk (composite) variability for H$_2$O (S/N$_{\rm{max}}$ = 14.0), CO (S/N$_{\rm{max}}$ = 13.0), and CH$_4$ (S/N$_{\rm{max}}$ = 14.9) molecular absorptions bands (see Figure \ref{fig:Vertical_Var_Map}). We find modulation in the wavelengths associated with the suspected forsterite cloud layer to be anti-correlated with wavelengths corresponding to H$_{2}$O and CO and correlated with low-altitude CH$_{4}$ and upper atmospheric temperature (see \S \ref{ssec:unique_maps}, \S \ref{ssec:discussion:anticorr} and Figure~\ref{fig:Vertical_Var_Map}). We suggest that cloud formation drives the temperature and chemical structure in the overlying atmosphere, leading in part to oxygen sequestration and disequilibrium carbon chemistry in the upper atmosphere.

    \item We infer harmonic modes associated with the high-altitude, CH$_4$ rotational-vibrational $\nu_3$ band distinct from those found in lower-altitude CH$_4$ band wavelengths (see \S \ref{ssec:unique_maps}, \S \ref{ssec:Discussion:auroralheating} and Figure~\ref{fig:Vertical_Var_Map}). Although the periodicities in both inferred vertical variability maps are comparable, they are anti-correlated, with the lower-altitude presumed CH$_4$ absorption features transitioning to emission features. We interpret this emission as upper atmospheric heating (potentially of auroral origin) with a periodicity set by the CH$_{4}$ abundance deviations, which appear to be influenced by forsterite cloud modulation (as discussed in \S \ref{ssec:discussion:anticorr}).
    
\end{enumerate}

Follow-on, multi-rotational and time-resolved spectroscopy of SIMP~J0136 and other highly variable brown dwarfs and directly-imaged gas giant exoplanets will be necessary to draw large-scale conclusions on the connection between cloud formation, atmospheric chemistry, and temperature structure in substellar objects.

\section*{Acknowledgements}

\par M.K.P., F.P.C., and P.A.K. would like to thank the United States Air Force Academy Department of Physics and Meteorology, Martinson Honors Program, Space Physics \& Atmospheric Research Center, and Center for Space Situational Awareness Research for supporting and enabling this research. Furthermore, we would like to thank the anonymous reviewer for their thoughtful and detailed feedback and suggestions. 

\par \'E.A., R.D., and B.B. acknowledge the financial support of the FRQ-NT through the Centre de recherche en astrophysique du Québec as well as the support from the Trottier Family Foundation and the Trottier Institute for Research on Exoplanets.

A.M.M acknowledges support from the National Science Foundation Graduate Research Fellowship Program under Grant No. DGE-1840990.

J. M. V acknowledges support from European Research Council Starting Grant Exo-PEA (Grant agreement No. [101164652]).

This work is based on observations made with the NASA/ESA/CSA James Webb Space Telescope. The data were obtained from the Mikulski Archive for Space Telescopes at the Space Telescope Science Institute, which is operated by the Association of Universities for Research in Astronomy, Inc., under NASA contract NAS $5-03127$ for JWST. These observations are associated with programs GT 1209 (PI: Artigau) and GO 3548 (PI: Vos). The data described here may be obtained from the MAST archive at \dataset[doi: 10.17909/zryr-vs58]{https://doi.org/10.17909/zryr-vs58} and 
\dataset[doi:10.17909/T9RP4V]{https://dx.doi.org/10.17909/T9RP4V}.

\par Approved for unlimited public release (United States Air Force, Public Affairs \# USAFA-DF-2025-879). The views expressed in this article, book, or presentation are those of the author and do not necessarily reflect the official policy or position of the United States Air Force Academy, the Air Force, the Department of Defense, or the U.S. Government.

\software{Imber \citep{Imber:2023,Imber:2024}, Astropy \citep{astropy:2013, astropy:2018,Astropy2022}, Dynesty \citep{Speagle2020}, Lightkurve \citep{Lightkurve:2018}, Matplotlib \citep{Matplotlib}, Pandas \citep{Pandas}, Picaso \citep{Picaso}, Scipy \citep{scipy2020}}

\clearpage
\bibliographystyle{aasjournal}
\bibliography{main,additional}




    

\end{document}